# Excitonic transport driven by repulsive dipolar interaction in a van der Waals heterostructure

Zhe Sun[1,2*], Alberto Ciarrocchi[1,2], Fedele Tagarelli[1,2], Juan Francisco Gonzalez Marin[1,2], Kenji Watanabe[3], Takashi Taniguchi[4], Andras Kis[1,2*]

[1]*Institute of Electrical and Micro Engineering, École Polytechnique Fédérale de Lausanne (EPFL), CH-1015 Lausanne, Switzerland*
[2]*Institute of Materials Science and Engineering, École Polytechnique Fédérale de Lausanne (EPFL), CH-1015 Lausanne, Switzerland*
[3]*Research Center for Functional Materials, National Institute for Materials Science, 1-1 Namiki, Tsukuba 305-0044, Japan*
[4]*International Center for Materials Nanoarchitectonics, National Institute for Materials Science, 1-1 Namiki, Tsukuba 305-0044, Japan*

*\*Correspondence should be addressed to: Zhe Sun (zhe.sun@epfl.ch) and Andras Kis (andras.kis@epfl.ch)*

**Dipolar bosonic gases are currently the focus of intensive research due to their interesting many-body physics in the quantum regime. Their experimental embodiments range from Rydberg atoms to GaAs double quantum wells and van der Waals heterostructures built from transition metal dichalcogenides. Although quantum gases are very dilute, mutual interactions between particles could lead to exotic many-body phenomena such as Bose-Einstein condensation and high-temperature superfluidity. Here, we report the effect of repulsive dipolar interactions on the dynamics of interlayer excitons in the dilute regime. By using spatial and time-resolved photoluminescence imaging, we observe the dynamics of exciton transport, enabling a direct estimation of the exciton mobility. The presence of interactions significantly modifies the diffusive transport of excitons, effectively acting as a source of drift force and enhancing the diffusion coefficient by one order of magnitude. The repulsive dipolar interactions combined with the electrical control of interlayer excitons opens up appealing new perspectives for excitonic devices.**

Van der Waals (vdW) heterostructures with a type-II band alignment enable the formation of long-lived interlayer excitons (IXs) composed of charges that are spatially separated in distinct layers[1]. In contrast to intralayer excitons in monolayer transition metal dichalcogenides (TMDCs), the spatial separation of charges gives rise to a sizable permanent out-of-plane electrical dipole moment, which makes IXs a promising platform to realize electrically controlled excitonic devices[2]. These are a new class of solid-state devices[3,4] for information and signal processing, analogous to electronic or spintronic devices, but based on encoding information in the amplitude and/or pseudo-spin of exciton currents, which can be controlled using electrical fields. As a result, exciton transport in vdW heterostructures has recently attracted a growing interest[5–8]. Many basic questions however remain open, such as the nature of mutual interactions between IXs. Moreover, exciton transport in vdW heterostructures has so far been attributed only to exciton diffusion currents and the observation of long IX transport distances is usually assigned to a large effective diffusion coefficient, a phenomenological parameter that conceals the role of the repulsive dipolar interaction on IX transport. In analogy with semiconducting devices, it would also be advantageous to introduce exciton drift currents in order to increase the range of exciton motion in a device and enhance control over exciton transport.

Here, by imaging the temporal evolution of IX cloud, we reveal repulsive dipolar exciton-exciton interactions as the driving force behind IX transport, acting as an effective source of a drift field. Concurrently, we deduce the exciton mobility directly from power-dependent drift velocities. Our findings, combined with the electrical control of IXs, pave the way to controlling the motion of IXs over long distances.

In our work, instead of investigating IXs in a heterobilayer, we introduce a monolayer hBN (1L-hBN) as a thin spacer between monolayers of WSe$_2$ (1L-WSe$_2$) and MoSe$_2$ (1L-MoSe$_2$). The motivations are twofold: firstly, the spacer can increase the separation between the electrons and the holes, enhancing the size of the electrical dipole by a factor of ~1.5[5]. Secondly, the presence of a moiré potential with an amplitude on the order of 100 to 200 meV[9–11] in MoSe$_2$/WSe$_2$ heterostructures localizes IXs[12] by effectively decreasing the diffusion coefficient[7,13]. The spacer weakens the moiré potential and reduces its period due to the large lattice mismatch between hBN and MoSe$_2$/WSe$_2$ while retaining a sufficiently strong transition dipole moment for hosting bright IXs[14].

Figures 1a and 1b show an optical image and a schematic of the device. It consists of a WSe$_2$/hBN/MoSe$_2$ heterotrilayer encapsulated in hBN, with a transparent global top gate and several local back gates. Multiple local back gates allow us to control the exciton flux by applying a laterally modulated vertical electric field (Figure 1b). The yellow-shaded area in Figure 1a indicates the heterotrilayer region where we perform the spatial and time-resolved photoluminescence (PL) measurements at 4.6 K. We use a sub-picosecond 725 nm pulsed laser with an 80 MHz repetition rate to excite the left part of the heterostructure and to generate an initial population of IXs (red spot in Figure 1a). Figure 1c presents the PL intensity of IXs at the excitation spot as a function of emission wavelength and average laser power $P_{ave}$. In Figure 1d we show the peak intensity and peak energy extracted from Figure 1c. With increasing laser power, the peak intensity first increases linearly and then begins to saturate at high laser powers (> 250 μW).

The observed blue-shift of IX energy with increasing laser power[5,15] is due to the repulsive exciton-exciton interaction, which can be decomposed into two terms: the dipolar repulsion, which is valley-independent, and the exchange interaction, which is determined by the valley indices[16]. The dipolar interaction is purely repulsive and can be estimated using a parallel-plate capacitor model[17–19]. This gives a lower bound on the exciton density, as it does not account for a reduction in interaction energy due to a rearrangement of the interlayer excitons, caused by Coulombic repulsion. Following the rearrangement and reduction in interaction energy, a larger density of IXs will be required to achieve the same energy shift. The exchange interaction has a more complex dependence on the electron-hole separation, since it can change from repulsive to attractive. As the vertical separation of electrons and holes becomes larger, the exchange interaction decreases and becomes negative when the vertical separation is larger than the Bohr radius of IX $a_0$ [19,20]. In our case, since the separation between the 1L-WSe$_2$ and 1L-MoSe$_2$ is around 0.9 nm and similar to the Bohr radius of IX ($a_0 \sim 1$ nm), we can neglect the exchange interaction and only consider the dipole-dipole interaction.

In order to quantify the influence of dipole-dipole interactions on exciton transport using time-resolved imaging, we first estimate the exciton initial density $n_0$ from the blue-shift $\delta E_{XX}$, via the parallel plate capacitor model [17–19]:

$$\delta E_{XX} = n_0 U_{XX} = n_0 \frac{e^2 d}{\varepsilon_0 \varepsilon_{HS}}, \qquad (1)$$

where $d = 0.9$ nm is the out-of-plane dipole size of IXs, $\varepsilon_0$ is the vacuum permittivity, $\varepsilon_{HS} = 6.26$ is the effective relative permittivity of the WSe$_2$/hBN/MoSe$_2$ heterotrilayer, and $U_{XX} \sim 2.6$ µeV·µm$^2$ is the exciton-exciton interaction strength (see Supplementary Note 1). We deduce an exciton density at $P_{ave}$ = 50, 100, 200 µW of about $2 \times 10^{11}$, $4 \times 10^{11}$, $8 \times 10^{11}$ cm$^{-2}$ respectively. The exciton densities we extract from the spectral shift are consistent with the values estimated from the applied laser powers (see Supplementary Note 2).

To image the spatial and temporal distribution of IXs, we use the setup depicted in Figure S3, in which the emitted photons are filtered (< 1.45 eV) and sent to either a charge-coupled device (CCD) camera or to a homemade scanning avalanche photodiode (APD) system (see Supplementary Note 3). Figures 1f and g show CCD images of the normalized PL emission intensity from IXs, acquired for different excitation powers. Compared with the CCD image of the focused excitation spot (Figure 1e), the spatial profile of IXs extends farther and exhibits a growing size with increasing excitation power, signaling the presence of strong repulsive exciton-exciton interactions.

To acquire the map of PL intensity as a function of time and position $I(x, y, t)$, the APD is scanned in the image plane across the emission spot[21,22]. We use a time-correlated photon counting module (TCPCM) to record the photon clicks. In order to rule out the decay of PL intensity induced by radiative emission, the raw data $I(x, y, t)$, proportional to the exciton density distribution $n(x, y, t)$, are normalized at each recorded time $t_i$ to obtain $I_{norm}^{time}(x, y, t_i) = I(x, y, t_i)/max(I(x, y, t_i))$ (see Supplementary Note 4). We present 2D spatial profiles of IXs at different times in Figure 2a for $P_{ave}$ = 200 µW. To further analyze the expansion of the spatial profile, the area of IXs at each time is extracted by counting the number of pixels for which the normalized PL intensity $I_{norm}^{time}(x, y, t_i)$ is higher than 0.2. This is shown in Figure 2c for three different excitation powers. In the absence of spatial constraints[23,24], the exciton cloud is expected to grow linearly as a function of time, with the evolution of the exciton density $n(x, y, t)$ due to diffusion described by $\frac{\partial n}{\partial t} = D\nabla^2 n - \frac{n}{\tau}$, where $D$ denotes the diffusion coefficient and $\tau$ the exciton radiative lifetime.

We find however that the area occupied by the IX cloud initially ($t$ < 2 ns) grows at a higher, power-dependent speed, but then slows down to a speed which is independent of the excitation power and initial exciton density. Even though in our case the motion of IXs is limited by the finite size of the heterostructure, this cannot explain the significant power dependency at early times, when the expansion is not constricted by the edges. Instead, we attribute the observed exciton cloud dynamics to dipolar interactions.

A sublinear increase of the IX cloud area with time has been observed before[13,25] and has been attributed to two possible mechanisms. First is the effect of the strong moiré potential introducing a modification of diffusivity $D = D_0 e^{-\frac{U_{moire}}{nU_{XX}+k_BT}}$, where $D_0$ denotes the bare diffusivity and $U_{moire}$ the depth of the moiré trapping potential[13]. The second is the generation of electron-hole plasma at exciton densities exceeding the Mott transition density $n_{Mott}$ ~ $10^{13}$

cm$^{-2}$ [25,26]. Neither of these reports however show a large area of transport (> 5 μm$^2$) in dilute excitonic gases ($n_0 \leqslant n_{Mott}$). We emphasize that in our work, in contrast with previous results, we observe a significant power-dependent expansion of the IX cloud in the dilute regime. In addition, in the heterotrilayer, the 1L-hBN spacer between WSe$_2$ and MoSe$_2$ is expected to weaken the trapping due to the moiré potential by increasing the spatial separation between the electron and hole wavefunctions, thereby facilitating the propagation of IXs[7]. Numerical simulations that include the effect of the moiré potential fail to reproduce our data (Supplementary Note 5). We therefore explain exciton transport using repulsive dipolar interactions only, decreasing in strength as the IXs cloud expands and decays radiatively.

To identify the contribution of exciton-exciton interactions, we introduce a power-related term into a 2D drift-diffusion equation and solve it numerically (see Supplementary Note 6). The drift-diffusion equation describes the spatial and temporal distribution of exciton density $n(x, y, t)$:

$$\frac{\partial n}{\partial t} = D \nabla^2 n + \frac{\mu}{e} \nabla (n \nabla \delta E) - \frac{n}{\tau} \qquad (2)$$

where $\mu$ is the exciton mobility which can be expressed using $D$ and temperature $T$ via Einstein relation $\mu = De/(k_B T)$ and $\delta E$ is the total potential energy of IXs. The first term on the right-hand side of equation (2) is the diffusion term, while the second term denotes the drift term. Here, $\delta E = \delta E_{XX} = n(x, y, t) \cdot U_{XX}$, leading to an exciton potential energy which varies both in time and space. The excitation power enters the equation via the initial exciton density $n_0(x, y)$, which is expected to be of the same order of magnitude as the exciton density determined from the interaction-induced energy shift (Figure 1d). We determine the lifetime of IXs to be about $\tau = 3.5$ ns from time-resolved PL measurements (Supplementary Note 7).

To highlight the effect of the drift term on exciton transport, we show on Figure 2b the simulated exciton area as a function of time with and without $\delta E_{XX}$ (solid and dashed lines). We use $U_{XX} = 2.6$ μeV·μm$^2$, $n_0 = 1 \times 10^{11}$ cm$^{-2}$ and do not take the finite size of the heterotrilayer into consideration. When $U_{XX}$ is neglected, the exciton area increases linearly with time and the slope is proportional to $D$. With $U_{XX}$ included, the area first increases sublinearly while at later times ($t > 2.5$ ns) when the drift term almost vanishes, the slope becomes the same as in the case of neglected $U_{XX}$. Similar simulations have successfully reproduced the experimental results of the indirect exciton diffusion in GaAs double quantum wells with strong dipole-dipole interactions[27]. We fit our data in Figure 2c using the same model and take the boundaries of the heterostructure into account (see Supplementary Note 8). To be consistent, we employ the same condition ($n_{norm}^{time}(x, y, t = t_0) > 0.2$) to calculate the exciton area. In Figure 2c, we use $D = 0.15$ cm²/s and $U_{XX} = 2.6$ μeV·μm² as parameters and treat the initial exciton density $n_0$ as a free variable to fit out data. The deduced exciton densities $n_0$ are similar to the values which we determined from the blue-shift energy and the applied laser power, confirming the consistency of our model. Considering the effective diffusion coefficient, this is enhanced by a factor of ~ 12 for $P_{ave} = 200$ μW when $t \leq 1$ ns (the ratio of slope between the red dashed line and the black dashed line).

We use the Einstein relation first to estimate the exciton mobility from the diffusion constant, finding $\mu \sim 380$ cm²/(V·s), similar to the low-temperature mobility of single charge carriers in monolayer TMDCs[28–30]. This indicates that at this temperature, exciton transport could be limited by the same mechanism, namely charged impurity scattering.

We now turn towards the drift term and quantify the relationships between exciton drift velocities and mobility, in analogy with transport of charge carriers in semiconductors. To visualize this relationship, we use an alternative way to normalize the raw data $I(x, y, t)$ by normalizing at each spatial coordinate $(x_i, y_i)$ by $I_{norm}^{space}(x_i, y_i, t) = I(x_i, y_i, t)/max(I(x_i, y_i, t))$ (see Supplementary Note 4). This allows us to observe the spatial distribution of IX lifetime as well as the spatially and temporally resolved IX transport. By scanning the APD along $x = 0$ in Figure 2a with a finer step, we obtain one-dimensional (1D) normalized

PL data $I_{norm}^{time}(0, y, t)$ and $I_{norm}^{space}(0, y, t)$ for different excitation powers, as shown in Figures 3a, b and c. From $I_{norm}^{time}$, we observe that IXs propagate about $y_{drift} \sim 1.5$ μm in the $+y$ direction, as indicated by the white dashed lines in Figure 3a, b and c. The lower panels distinctly show the time delay at $y > 0$ generated during IX propagation in the $+y$ direction. Figure 3d shows the simulated normalized exciton density distribution $n_{norm}^{time}(0, y, t)$ and $n_{norm}^{space}(0, y, t)$ for $n_0 = 4 \times 10^{11}$ cm$^{-2}$ using equation (2). $I_{norm}^{space}$ allows us to extract the effective speed $v_{eff}$ of IXs at different powers (see Supplementary Note 9). The black dashed lines in the lower panels of Figure 3a, b and c are guides for the eye which highlight the effect of excitation power and initial exciton density on $v_{eff}$. We can further decompose $v_{eff}$ into its diffusion and drift components as $v_{eff} = v_{diff} + v_{drift}$. At a higher excitation power, due to the repulsive interactions, IXs experience a stronger effective drift field $F_{drift}$, which can be deduced from the spectral blue-shift $\delta E_{XX}$ and drift distance $y_{drift}$ via $F_{drift} = \delta E_{XX}/y_{drift}/e$. In analogy with the transport of charge carriers in an electric field, the neutral exciton mobility $\mu$ can be approximately expressed using $\mu = v_{drift}/F_{drift}$, allowing us to directly estimate the exciton mobility without using the Einstein relation[31]. We extract $v_{eff}$ at different powers from $I_{norm}^{space}(0, y, t)$ and plot them with the spectral blue-shift $\delta E_{XX}$ in Figure 3e. The error bars of $v_{eff}$ are given by the linear fits along the black dashed lines. By applying a linear fit to the data in Figure 3e, we estimate $\mu \sim 440$ cm$^2$/(V·s) which is consistent with the value calculated from the diffusion coefficient. The presence of the power-dependent excitonic drift force shows that the repulsive dipolar interactions can be used to control IX transport.

Next, we present how combining the repulsive interactions together with an external electric field can be used to control the motion of IXs. We generate a laterally modulated electric field $F_{el}(x)$ along the $x$ direction, which creates a spatially varied but time-independent energy profile acting on IXs: $\delta E_{el} = -ed \cdot F_{el}(x)$. The upper panels in Figure 4a and b show the schematics of the energy profiles as well as the expected exciton motion for a back-gate voltage $V_{bg} = -2$ V and 2 V. The middle panels in Figure 4a and b present CCD images of the normalized IX PL intensity using $P_{ave} = 200$ µW. The region enclosed by the black dashed lines indicates the position of the local back gate. By tuning the gate region higher or lower in energy with respect to its surroundings, we generate a potential barrier or a trap, effectively controlling the propagation distance of IXs along the $+x$ direction. We measure the 1D normalized PL intensity $I_{norm}^{time}(x, 0, t)$ along $y = 0$ using the scanning APD system. As presented in the lower panels in Figure 4a and b, we clearly observe the process of IX gas moving into the back-gate region as we adjust the electro-static potential configuration from a barrier to a trap. Figure 4c shows instead the 1D normalized PL intensity $I_{norm}^{time}(x, 0, t)$ for $V_{bg} = -1$ V, i.e. a barrier configuration of lower amplitude with respect to Figure 4a. Here the excitons first flow towards the gate region of lower potential, but then surprisingly move away from the gate region after ~ 1 ns. This phenomenon is another manifestation of the repulsive exciton-exciton interaction, and its interplay with the electrostatic potential in the device. In order to explain this observation, in Figure 4e we show a schematic of the energy profile of the dipolar repulsion $\delta E_{XX}(t) = n(t) \cdot U_{XX}$ and the electro-static energy $\delta E_{el}$. Initially, when the exciton density $n_0$ is high, $\delta E_{XX}$ dominates over $\delta E_{el}$ such that the exciton flux is pushed towards the gate region. As the exciton density decreases due to the spatial expansion and the radiative emission, $\delta E_{el}$ becomes higher than $\delta E_{XX}$, resulting in the exciton flux leaving the gate region. The observation of this competitive phenomenon requires that $n_0 U_{XX}$ is larger than but comparable to $\delta E_{el}$. In this measurement, we create such condition by using $P_{ave} = 200$ µW, which corresponds to $\delta E_{XX}(t = 0) = \delta E_{XX}^0 \sim 20$ meV, and $\delta E_{el} \sim 9$ meV for $V_{bg} = -1$ V (see Supplementary Note 1). We introduce $\delta E_{el}$ into the simulation to better evaluate the interplay between $\delta E_{XX}$ and $\delta E_{el}$. Here we assume that the electro-static potential has a form of an harmonic potential $\delta E_{el} \propto (x - x_c)^2$ with height $\delta E_{el}^0$ (insert in Figure 4f). Figure 4f shows the experimental results as well as the simulations of the propagation distance $L_x$ along the $+x$ direction as a function of $\delta E_{el}^0/\delta E_{XX}^0$ (see Supplementary Note 10). Therefore, adjusting the ratio $\delta E_{el}^0/\delta E_{XX}^0$ allows us to precisely control the propagation distance of IXs.

Our results demonstrate that the repulsive dipolar interactions in dilute excitonic gases have a strong influence on exciton transport and that they act as a source of a drift force. Time-resolved PL imaging enables us to visualize the dynamic evolution of IXs in the WSe$_2$/hBN/MoSe$_2$ heterotrilayer and allows us to quantify the diffusion coefficient and exciton mobility which play a central role in the prospect for applications of excitonic devices. Our findings constitute a crucial step towards using spatial patterns of laser field to control the propagation of IXs[32,33]. The excitons that are driven away from the laser hot spot by the repulsive exciton-exciton interactions are expected to constitute a cold excitonic gas. Many exotic phases of matter and emergent phenomena might appear in it, including Bose Einstein condensation and high-temperature superfluidity[34–36]. Finally, a strong spatial confinement of IXs that experience strong repulsive dipolar interactions leads to nonlinearities in energy[37]. Due to the additional 1L-hBN spacer and thus stronger repulsive interactions, we expect the nonlinearity of localized excitons in the heterotrilayer to be more significant.

**ACKNOWLEDGEMENTS**
We acknowledge many helpful discussions with A. Delteil as well as D. Unuchek and A. Avsar. The authors acknowledge the help of Zdenek Benes of CMi for electron beam lithography. This work was financially supported by the European Research Council (grant no. 682332), the Swiss National Science Foundation (grants no. 175822, 177007 and 164015). This project has received funding from the European Union's Horizon 2020 research and innovation programme under grant agreement No 785219 and 881603 (Graphene Flagship Core 2 and Core 3 phases) as well as support from the CCMX Materials Challenge grant "Large area growth of 2D materials for device integration". K.W. and T.T. acknowledge support from the Elemental Strategy Initiative conducted by the MEXT, Japan (Grant Number JPMXP0112101001) and JSPS KAKENHI (Grant Numbers 19H05790 and JP20H00354).

**AUTHOR CONTRIBUTIONS**

Z.S. built the experimental setups, performed the optical measurements and analysed the data with input from A.K. A.K. initiated and supervised the project. A.C. fabricated the device. F.T. worked on device fabrication. J.F.G.M. contributed to the initial stages of setup development. K.W. and T.T. grew the h-BN crystals. Z.S. and A.K. wrote the manuscript with input from all authors.

**COMPETING FINANCIAL INTERESTS**

The authors declare no competing financial interests.

**DATA AVAILABILITY**

The data that support the findings of this study are available from the corresponding author on reasonable request.

**METHODS**

**Device fabrication**

Thin Cr/Pt (2-3 nm) local back gates were patterned using electron-beam lithography and deposited on a silicon substrate using electron-beam evaporation. The heterostructure was fabricated using a polymer-assisted transfer method. Flakes were first exfoliated on a polymer double layer. After monolayers were optically identified by photoluminescence, the bottom layer was dissolved with a solvent and free-floating films with flakes were obtained. These were transferred using a home-built setup with micromanipulators to carefully align flakes on top of each other. Polymer residue was removed with a hot acetone bath. Afterwards, the heterostructure was thermally annealed under high-vacuum conditions ($10^{-6}$ mbar) for 6 h at 340 °C. Finally, electrical contacts (80 nm Pd for the contacts to the flakes, 8 nm Pt for the global top gate) were patterned using electron-beam lithography and deposited using electron-beam evaporation.

**Time-resolved optical characterization**

A confocal microscope is used to optically excite IXs and collect the emitted photons through the same objective with a working distance 4.5 mm and NA = 0.65. IXs are excited using sub-ps pulses with a repetition rate of 80 MHz from Ti:Sapphire laser. The collected photons are sent to a APD (Excelitas Technologies, SPCM-AQRH-16) mounted on a 2D motorized translational stage. The output of the APD is connected to a time-correlated photon counting module (TCPCM) with a resolution of 12 ps r.m.s. (PicoQuant, PicoHarp 300), which measures the arrival time of each photon. Technical details can be found in Supplementary Note 3.

**Numerical simulations**

The 2D drift-diffusion equation are solved numerically using Forward Time Centered Space (FTCS) method. The method is based on central difference in space and the forward Euler method in time. Details can be seen in Supplementary Note 6. Numerical simulations that include the effect of moiré potential is performed using the same method and are summarized in Supplementary Note 5.

**FIGURES**

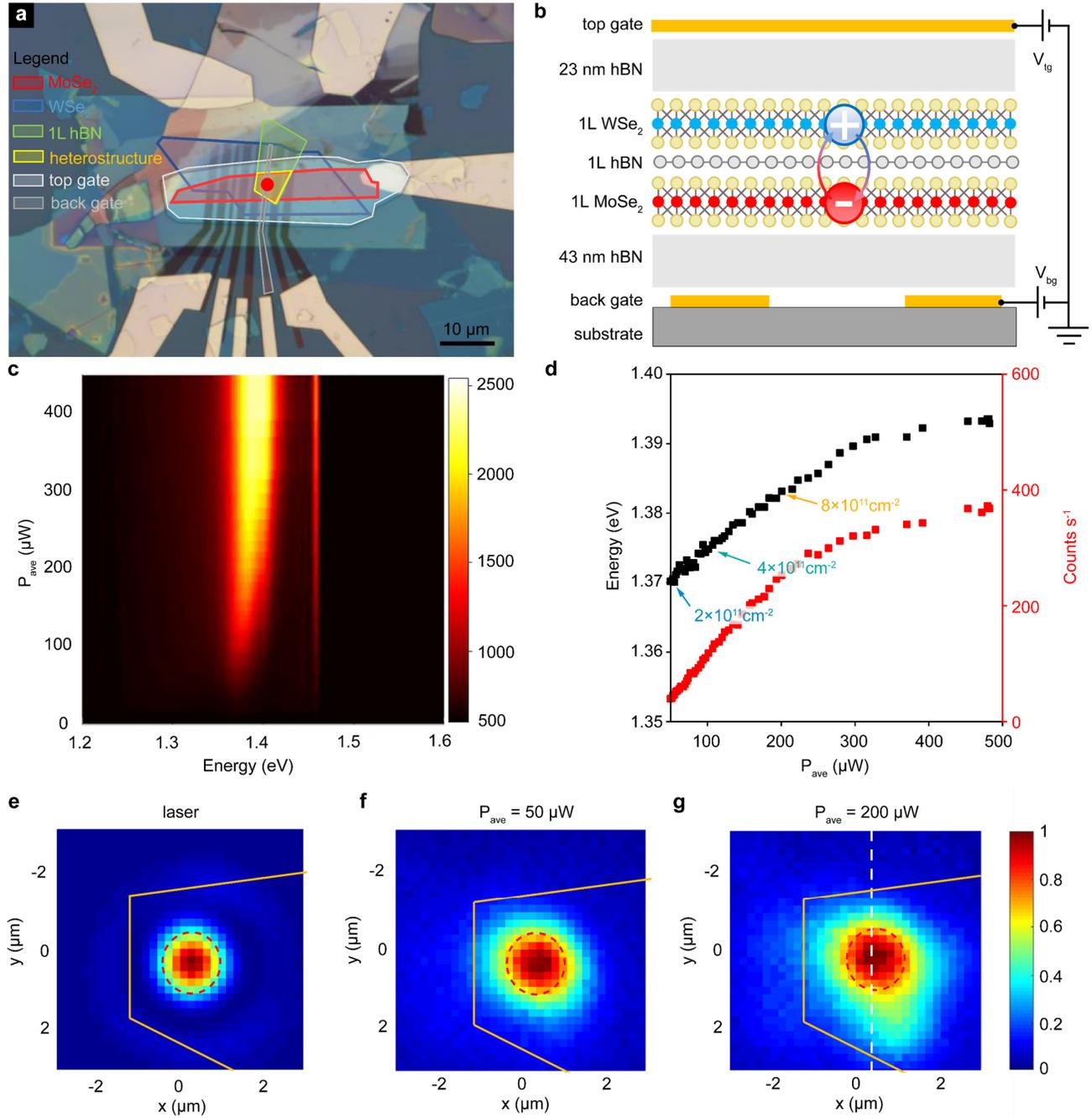

Figure 1. **a,** Optical image of the device, highlighting the regions for different materials. Scale bar, 10 μm. **b**, Schematic of the encapsulated heterotrilayer device with the global top and bottom split gates, as well as electrical connections. **c**, IX PL emission intensity as a function of laser average power $P_{ave}$ and energy **d**, PL peak intensity (red squares) and peak energy (black squares) extracted from c. The arrows point out the estimated exciton density at three different powers which will be used in the data in Figure 2. **e**, CCD image of the excitation laser spot. **f** & **g**, CCD images of the normalized IX PL intensity, acquired for $P_{ave}$ = 50 μW and 200 μW. The yellow solid lines indicate the shape of the heterostructure.

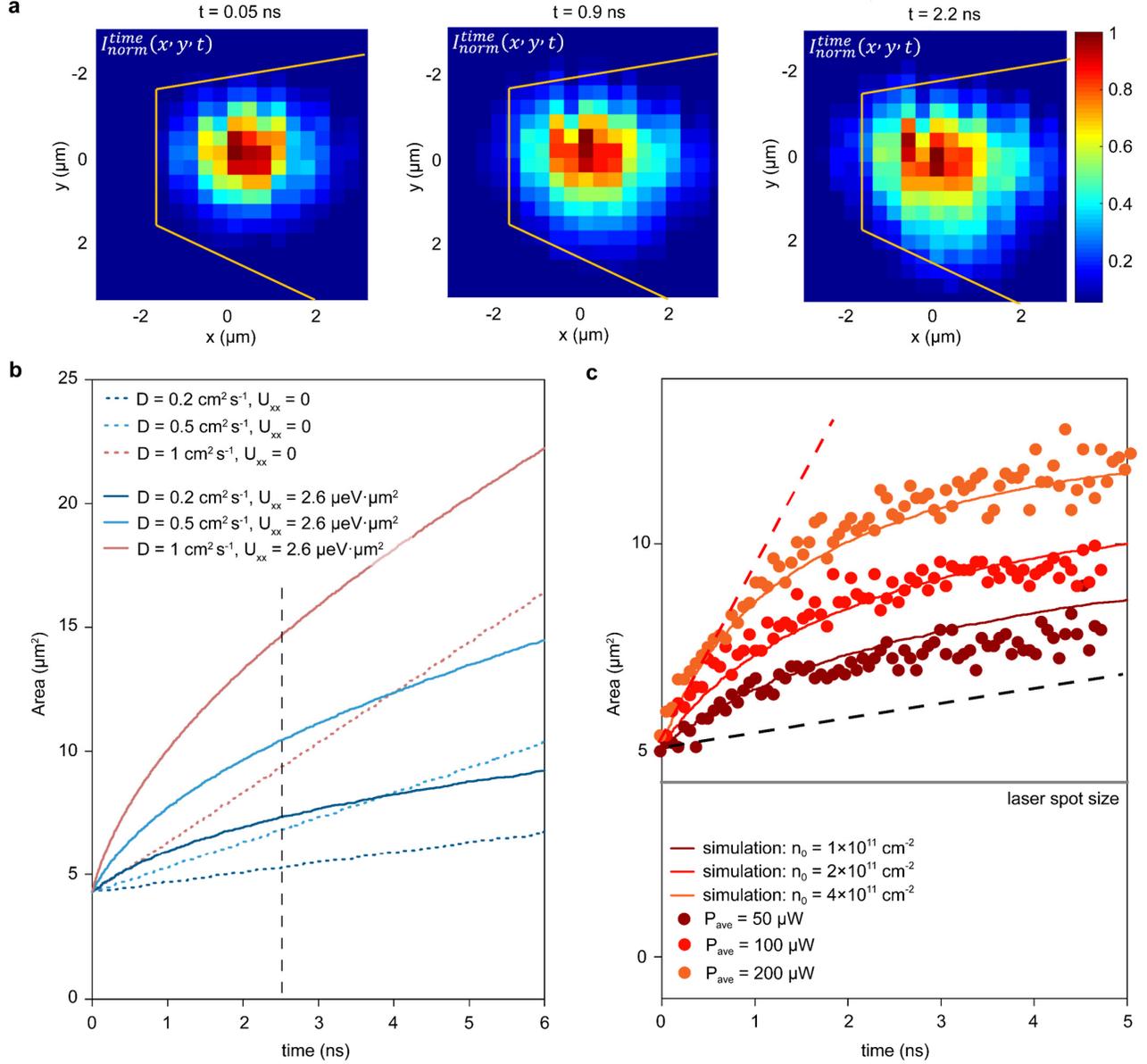

Figure 2. **a**, 2D PL images $I^{time}_{norm}$ for $P_{ave}$ = 200 μW acquired by the scanning APD system at different times after the excitation. The yellow solid lines indicate the shape of the heterostructure. **b**, Simulated exciton area as a function of time for different diffusion coefficients $D$. The fitting parameters are as follows. Thin solid lines: $U_{XX}$ = 0, $\tau$ = 3.5 ns; thick solid lines: $U_{XX}$ = 2.6 μeV·μm², $n_0 = 1 \times 10^{11}$ cm⁻², $\tau$ = 3.5 ns. **c**, Exciton area as a function of time for different excitation powers. The gray line shows the area of the excitation laser spot size (see Figure S5). The solid lines are fits using equation (2) for $n_0 = 1 \times 10^{11}$, $2 \times 10^{11}$ and $4 \times 10^{11}$ cm⁻² respectively. The fitting parameters are $U_{XX}$ = 2.6 μeV·μm², $\tau$ = 3.5 ns, $D$ = 0.15 cm²/s and $T$ = 4.6 K. The black dashed line indicates the simulated exciton area without considering $U_{XX}$. The red dashed line indicates the slope of the area increasing for $P_{ave}$ = 200 μW at $t$ < 1 ns.

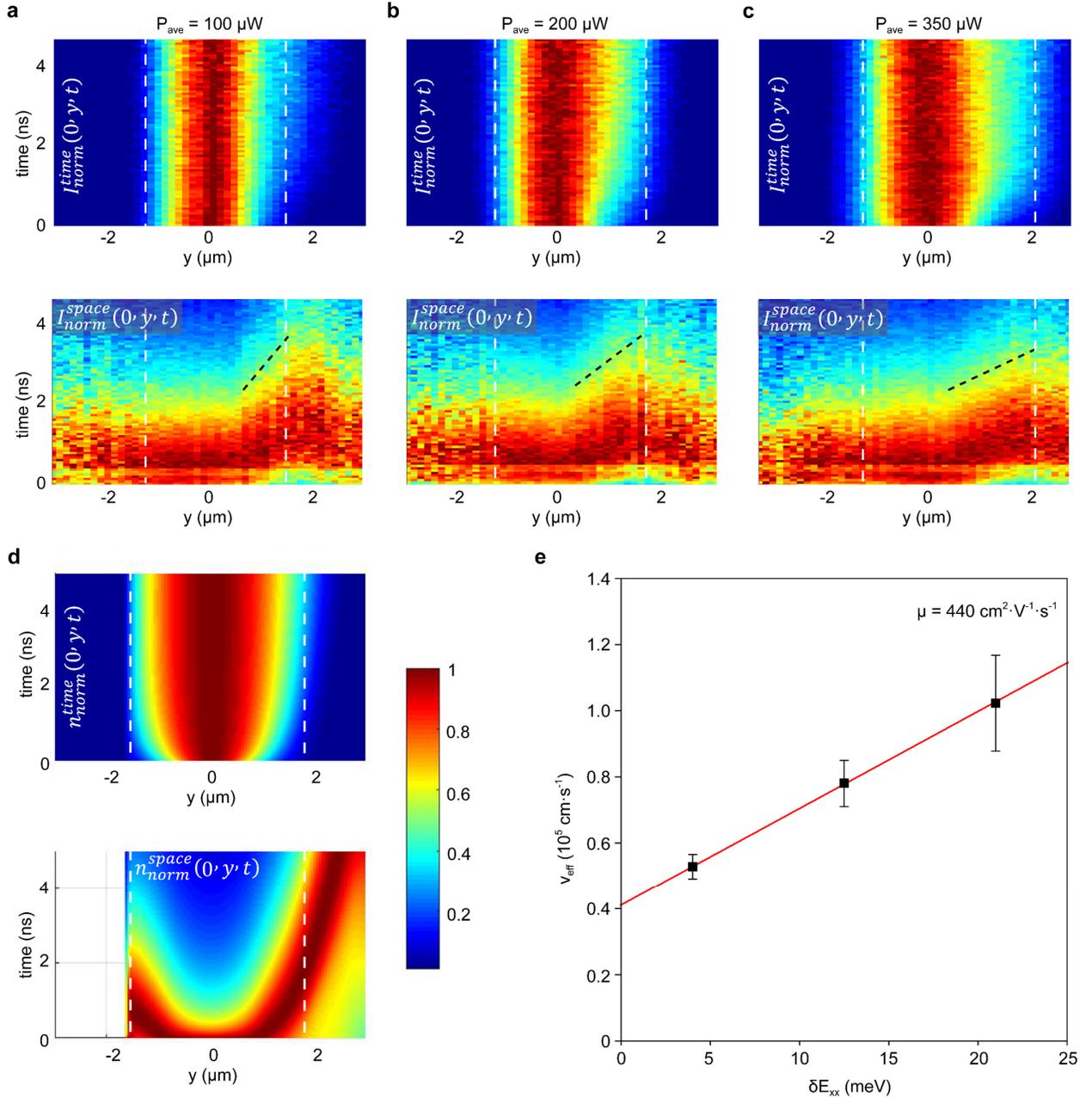

Figure 3. **a, b** and **c**, 1D normalized PL intensity along $x = 0$ in Figure 2a for $P_{ave}$ = 100 μW, 200 μW and 350 μW. Upper panel: $I_{norm}^{time}$; lower panel: $I_{norm}^{space}$. White dashed lines enclose the region of IXs transport. Black dashed lines are guides for the eye for $v_{eff}$. **d**, Simulation of 1D normalized exciton distribution along $x = 0$ using equation (2) for $n_0 = 4 \times 10^{11}$ cm$^{-2}$. Upper panel: $n_{norm}^{time}$; lower panel: $n_{norm}^{space}$. **e**, Exciton effective velocity $v_{eff}$ as a function of the spectral blue-shift $\delta E_{XX}$ extracted from Figure 1d. The error bars of $v_{eff}$ are given by the linear fits along the black dashed lines in a, b and c (see Figure S11). The red solid line is a linear fit to the data.

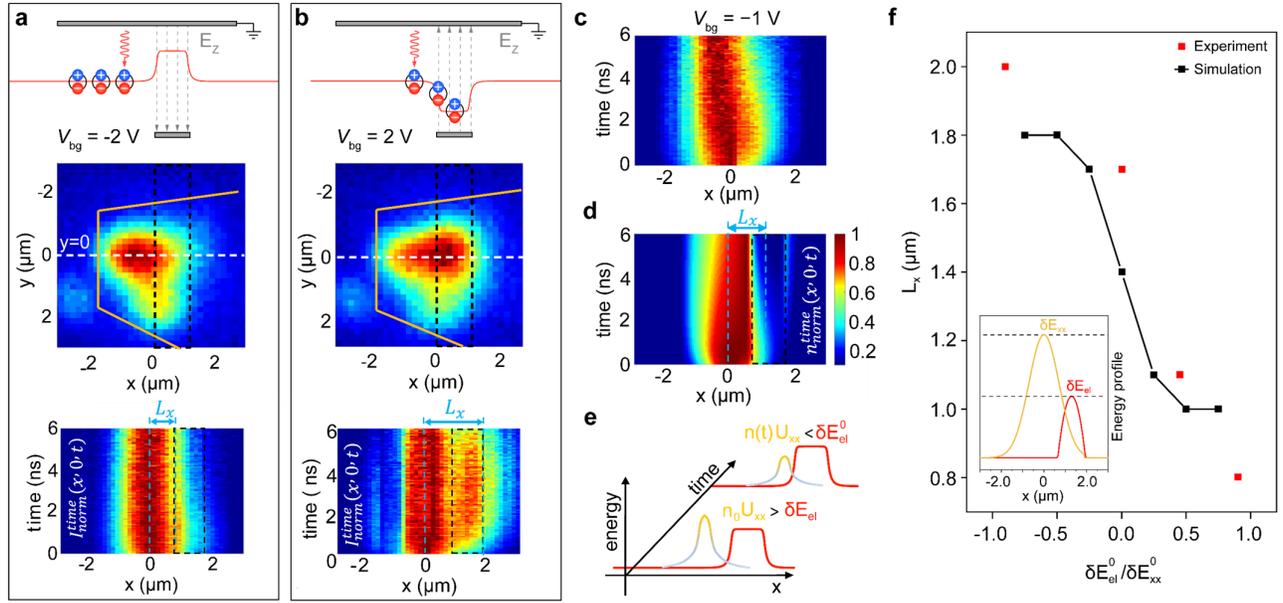

Figure 4. **a, b**, Effect of electro-static potential applied to the back-gate ($V_{bg}$) on the exciton spatial and temporal distribution. Upper panel: schematic of the energy profiles as well as the expected exciton motion; middle panel: CCD images of the normalized IX PL intensity; lower panel: 1D normalized PL intensity $I_{norm}^{time}$ along $y = 0$. The black dashed lines enclose the region of the local back gate. The yellow solid lines indicate the shape of heterostructure. **c**, 1D normalized PL intensity $I_{norm}^{time}$ along $y = 0$ for $V_{bg}$ = -1 V. **d**, Simulation of 1D normalized exciton distribution $n_{norm}^{time}$ along $y = 0$ for $\delta E_{el}^0/\delta E_{XX}^0 \sim 0.5$. **e**, Schematic of the energy profile of $\delta E_{XX}(t) = n(t)U_{XX}$ and $\delta E_{el}$. **f**, Exciton propagation distance $L_x$ as a function of $\delta E_{el}^0/\delta E_{XX}^0$. Insert: schematic of the energy profile in simulations. Red: $\delta E_{el}$; yellow: $\delta E_{XX}(t = 0)$.

18